\documentclass{article}
\usepackage{amsmath,amssymb,hyperref}

\newcommand{\mathd}{\mathrm{d}}
\newcommand{\tmem}[1]{{\em #1\/}}
\newcommand{\tmop}[1]{\ensuremath{\operatorname{#1}}}
\newcommand{\tmstrong}[1]{\textbf{#1}}
\newcommand{\tmtextbf}[1]{{\bfseries{#1}}}
\newcommand{\tmtextit}[1]{{\itshape{#1}}}
\newcommand{\tmtexttt}[1]{{\ttfamily{#1}}}
\begin{document}

\begin{center}
  {\tmstrong{\\
  Spontaneous Lorentz Violation, Negative Energy and the Second Law of
  Thermodynamics}}
\end{center}

\begin{center}
  Brian Feldstein
  
  {\tmem{Department of Physics}}

  {\tmem{Boston University}}
  
  {\tmem{Boston, MA, 02215}}
\end{center}

\begin{center}
  ABSTRACT
\end{center}

\paragraph{}We reconsider the possibility of violating the generalized second
law of thermodynamics in theories with spontaneous Lorentz violation. \ It has
been proposed that this may be accomplished in particular with a black hole
immersed in a ghost condensate background, which may be taken to break Lorentz
invariance without appreciably distorting the space-time geometry. \ In this
paper we show that there in fact exist solutions explicitly describing flow of
negative energy into these black holes, allowing for violation of the second
law in a very simple way. \ This second law violation is independent of any
additional assumptions such as couplings of the ghost condensate to secondary
fields, and suggests that violation of the null energy condition may be the
true source of \ pathology in these theories.

{\newpage}

\section{Introduction\label{intro}}

In {\cite{Sergeis}}, Dubovsky and Sibiryakov discussed a relationship between
theories with spontaneous violation of Lorentz invariance, and violation of
the generalized second law of thermodynamics (GSL) {\cite{Bekenstein}} in the
presence of black holes. \ That paper continues an intriguing line of inquiry
in which one attempts to place bounds on possibilities for low energy physics,
coming from what appear to be fundamental principles of quantum gravity. \
Other recent results in this direction have included an upper bound on
particle masses in theories with large numbers of species
{\cite{Dvali1,Dvali2}}, and a possible upper limit to the permitted size of
the fine structure constant {\cite{DaviesAlpha,BekensteinAlpha}}. \ See also
{\cite{entropybound,BoussoEntropy,gravityweak,viscosity1,viscosity2,IRobstruct}}.
\

\paragraph{}The specific setup discussed in {\cite{Sergeis}} involved a
``ghost condensate'' background surrounding a Schwarzschild black hole. \ The
ghost condensate is defined by a scalar field with an expectation value for
its kinetic term $\left\langle \nabla_{\mu} \varphi \nabla^{\mu} \varphi
\right\rangle > 0$, singling out a specific time-like direction for a breaking
of Lorentz symmetry {\cite{ghost,ghost2,ghost3,ghostinflate2,shinjiperturb}}.
\ With some assumed couplings of the ghost condensate to secondary fields, it
then becomes possible to construct a perpetual motion machine around the black
hole, violating the second law. \ In this paper, we will re-examine this
setup, the details of which will be reviewed in section (\ref{GSL}). \ We will
note that, in order to conclude that the GSL is actually violated, one must
carefully examine the energy flowing into the black hole in the ghost
condensate background under consideration. \ Although naively this energy flow
vanishes, we will point out that even seemingly small corrections to this
statement, from various possible sources, could ruin the purported GSL
violation and must be checked carefully.

\paragraph{}With this motivation, in section (\ref{ghost}) we will turn to an
examination of possible ghost condensate flows into black holes. \ Using a
perturbative approach, we will confirm the numerical result of
{\cite{Frolov}} that, contrary to the expectation from fluid dynamics, there
is a one parameter family of static spherically symmetric ghost condensate
flows, rather than a unique one. \ We will discuss the range of parameters for
which these flow solutions may be trusted, including possible effects from
higher derivative operators.

\paragraph{}In section (\ref{energy}), we will calculate the energy-momentum
tensor for these solutions, and show that they describe regions of either
positive {\tmem{or}} negative energy flowing into the black holes. \ The
negative energy solutions allow one to violate the GSL in a very simple way,
independent of the couplings to secondary fields. \ We will also comment on
the timescale for an instability in these flows, and argue that they should be
sufficiently long lived to easily allow violation of the GSL.

\paragraph{}We will conclude in section (\ref{discussion}), including a
discussion of whether it is really violation of the null energy condition,
rather than Lorentz invariance, which causes the difficulties with
thermodynamics in these theories.

\section{Generalized Second Law Violation\label{GSL}}

\paragraph{}The ghost condensate may be defined by a Lagrangian taking a form
such as
\begin{equation}
  \mathcal{L}= \frac{1}{2 \Lambda^4} (X - \Lambda^4)^2, \label{L}
\end{equation}
where $X \equiv \nabla_{\mu} \varphi \nabla^{\mu} \varphi .$ \ The energy
momentum tensor is
\begin{equation}
  T_{\mu \nu} = \frac{2}{\Lambda^4} (X - \Lambda^4) \nabla_{\mu} \varphi
  \nabla_{\nu} \varphi - \frac{1}{2 \Lambda^4} (X - \Lambda^4)^2 g_{\mu \nu},
  \label{T}
\end{equation}
and the equation of motion is
\begin{equation}
  \nabla_{\mu} \left( \frac{2}{\Lambda^4} (X - \Lambda^4) \nabla^{\mu} \varphi
  \right) = 0. \label{eqnmot}
\end{equation}
It follows that there are zero energy solutions for which $X = \Lambda^4$, and
which therefore violate Lorentz invariance spontaneously. \ In flat space
these solutions may be taken to have the form $\varphi = \Lambda^2 t$, with
$t$ identifying the time-like direction singled out for a breaking of Lorentz
symmetry.

\paragraph{}It is important to note that these $X = \Lambda^4$ zero energy
solutions exist not only in flat space-time, but in general curved backgrounds
as well. \ In particular, we may set up such a configuration in a black hole
background without affecting the underlying geometry {\cite{shinji}}. \ The
net effect is that we may construct an ordinary looking black hole, except
that at each point there is a preferred direction as chosen by the ghost
condensate field.

\paragraph{}We write the Schwarzschild black hole geometry in the form
\begin{equation}
  \tmop{ds}^2 = (1 - \frac{2 M}{r}) d \tau^2 - 2 \sqrt{\frac{2 M}{r}} d \tau
  \tmop{dr} - \tmop{dr}^2 - r^2 d \Omega^2 . \label{metric1}
\end{equation}
Here M is the mass of the black hole, and we have chosen natural units such
that the Planck scale is $M_{\tmop{pl}} = 1.$ \ $r$ is the usual radial
coordinate of Schwarzschild space-time, such that spheres at radius $r$ have
area $4 \pi r^2$, and $d \Omega^2$ is the standard metric on the unit sphere.
\ The coordinate $\tau$ measures the proper time as seen by observers who
freely fall into the black hole from rest at infinity, with surfaces of
constant $\tau$ being orthogonal to the world lines of such observers. \
$\tau$ is related to the usual killing time $t$ of Schwarzschild space-time
via the relation
\begin{equation}
  \tau = t + 2 M \left( 2 \sqrt{\frac{r}{2 M}} + \log \left( \frac{\sqrt{r} -
  \sqrt{2 M}}{\sqrt{r} + \sqrt{2 M}} \right) \right), \label{tau}
\end{equation}
in terms of which the metric takes the standard form
\begin{equation}
  \tmop{ds}^2 = (1 - \frac{2 M}{r}) \tmop{dt}^2 - \frac{1}{1 - \frac{2 M}{r}}
  \tmop{dr}^2 - r^2 d \Omega^2 . \label{metric2}
\end{equation}
Recall that $t$ labels the time-symmetry direction of the space-time, and that
it becomes singular at the horizon ($r_h = 2 M)$, unlike $\tau$.

\paragraph{}In terms of $\tau$, the zero energy ghost condensate solution in
Schwarzschild space-time with $X = \Lambda^4$ takes an extremely simple form:
\begin{equation}
  \varphi_0 = \Lambda^2 \tau . \label{phi}
\end{equation}
Indeed, the direction chosen by $\varphi$ in this solution is simply that of
freely falling observers, with $\nabla^{\mu} \varphi_0$ following their
geodesics.

\paragraph{}The argument in {\cite{Sergeis}} now proceeds as follows: \ Since
the ghost condensate picks out a special time-direction, it is simple to
couple it to a secondary field, $\psi$, in such a way that the particles of
$\psi$ obtain a maximum speed which is different from the speed of light. \ We
might imagine for example a coupling of the form
\begin{equation}
  \mathcal{L} \supset \frac{1}{2} \nabla_{\mu} \psi \nabla^{\mu} \psi +
  \frac{\varepsilon}{2} ( \frac{\nabla^{\mu} \varphi}{\Lambda^2} \nabla_{\mu}
  \psi)^2, \label{coupling}
\end{equation}
which gives the $\psi$ particles a propagation speed of $v = \frac{1}{\sqrt{1
+ \varepsilon}}$. \ The key point is that this non-Lorentz invariant speed
implies that the $\psi$ particles have a horizon at a different radius from
the usual horizon, and in turn a temperature different from the usual black
hole temperature. \ The horizon radius and temperature for $\psi$ are given by
\begin{equation}
  r_{\psi} = r_h (1 + \varepsilon) \label{rpsi}
\end{equation}
and
\begin{equation}
  T_{\psi} = v^3 T, \label{Tpsi}
\end{equation}
where $T$ is the standard black hole temperature $T = 1 / 8 \pi M.$ \ As might
have been expected, particles with a slower speed have a larger horizon, and
smaller temperature.

\ Particles without such a coupling to the ghost condensate of course still
have the usual horizon radius and temperature, and in this way we obtain a
black hole with rather peculiar thermodynamic properties; in particular it is
immediately unclear which temperature one should think of as the ``real''
temperature (if any), or which horizon area one should associate with the
black hole entropy (if any). \ Indeed, the authors of {\cite{Sergeis}} went on
to show that one may violate the second law of thermodynamics in this setup
with an appropriate set of hot shells surrounding the black hole. \ In
particular, suppose we have two particle species $1$ and $2$ with speeds $v_1
< v_2$ and temperatures $T_1 < T_2$. \ Now imagine placing two shells around
the black hole with temperatures $T_A$ and $T_B$ satisfying
\begin{equation}
  T_1 < T_A < T_B < T_2 . \label{Tineq}
\end{equation}
Moreover, let us assume that shell $A$ only interacts with particles of type
$1$, and shell $B$ only interacts with particles of type $2$. \ If one chooses
the temperatures appropriately then we may satisfy the inequalities
(\ref{Tineq}), while simultaneously satisfying the condition that the total
energy flowing into the black hole is equal to zero. \ The net result of this
setup for the type $1$ particles is then a flow of energy
{\tmem{{\tmem{into}}}} the black hole {\tmem{{\tmem{from}}}} the low
temperature shell, while for the type $2$ particles it is a flow of energy
{\tmem{{\tmem{out}}}} of the black hole {\tmem{{\tmem{onto}}}} the high
temperature shell. \ Since no total energy is being lost by the shells, this
``machine'' then has the net effect of transferring energy from the cold shell
to the hot shell, lowering the entropy of matter outside the black hole
without increasing the black hole area. \ It is in this way that this setup
manages to violate the second law of thermodynamics.{\footnote{In
{\cite{Jacobson3}} the authors construct a process to violate the second law
in this theory via a purely classical method, not involving Hawking
radiation.}}

\paragraph{}Now, a question one might immediately ask is: \ How can it be that
backgrounds which violate Lorentz invariance allow for the existence of
perpetual motion machines? \ In particular, why isn't it possible to violate
the second law of thermodynamics by using a regular fluid such as water
flowing into a black hole? \ After all, water picks out a specific frame, and
the phonons in water have a maximum speed which differs from the speed of
light. \ One could imagine setting up the same sort of device as outlined
above, but with the slow particles replaced by the phonons in water. \ The
reason such a device would fail to violate the GSL of course is simple: \
While the machine runs, there is a net flux of energy in water molecules
flowing into the black hole, and this energy causes the horizon area to grow
at a rate which is sufficient to overcome any purported GSL violation.

\paragraph{}The point is that it is the special property of the ghost
condensate of carrying essentially no energy-momentum which results in the
difficulties with the laws of thermodynamics. \ For this reason, it is
important to check carefully the claim that one can maintain the condition
$T_{\mu \nu} = 0$ in the ghost condensate background. \ In particular, any
perturbation to the equation of motion (\ref{eqnmot}) has the potential to
result in some small but non-zero energy flow during the running of the
perpetual motion machine, and could negate any GSL violation.

\paragraph{}The first such perturbation one might worry about is that of
higher derivative operators. \ In fact if one expands the ghost condensate in
small perturbations $\pi$ about the $X = \Lambda^4$ background, the equation
of motion (\ref{eqnmot}) becomes at linear order
\begin{equation}
  \nabla_{\nu} \xi^{\nu} \xi^{\mu} \nabla_{\mu} \pi + \xi^{\mu} \xi^{\nu}
  \nabla_{\mu} \nabla_{\nu} \pi = 0, \label{curvedpi}
\end{equation}
with $\xi^{\mu} \equiv \nabla^{\mu} \varphi_0 / \Lambda^2$ being the
normalized gradient of the background. \ In flat space-time this wave equation
takes the form
\begin{equation}
  \ddot{\pi} = 0. \label{pidotdot}
\end{equation}
It is therefore imperative for stability of these fluctuations that one adds
to the original ghost condensate Lagrangian some higher derivative operators
{\cite{ghost}} such as for example{\footnote{Of course, such terms are
automatically expected to be present in our effective field theory since no
symmetry forbids them. \ We will assume $\alpha$ to be roughly of order 1.}}
\begin{equation}
  \mathcal{L} \supset - \frac{2 \alpha}{\Lambda^2} \Box \varphi \Box \varphi .
  \label{higherd}
\end{equation}
At energies much less than $\Lambda$, such a term modifies the $\pi$ equation
of motion to{\footnote{This form for the low energy $\pi$ equation of motion
is independent of the details of the structure of the higher derivative terms.
\ This is because it follows from (\ref{stablepi}) that time derivatives of
$\pi$ are generally suppressed compared to its spatial derivatives, and so
(\ref{stablepi}) automatically includes the dominant modification to
(\ref{pidotdot}).}}
\begin{equation}
  \ddot{\pi} + \frac{\alpha}{\Lambda^2} \vec{\nabla}^4 \pi = 0
  \label{stablepi} .
\end{equation}
It is therefore important to analyze the effect of higher derivative operators
on the energy-momentum carried by the ghost condensate into black holes. \

\paragraph{}A second potentially problematic perturbation to the ghost
condensate flow comes from the couplings (\ref{coupling}) being used to alter
the speeds of the type $1$ and type $2$ particles.  The non-zero fluxes of
these particles, a key component in the construction of {\cite{Sergeis}}'s
perpetual motion machine, will lead to a perturbation to the ghost condensate
solution, and any resulting energy flux could be important.{\footnote{The setup of reference
{\cite{Jacobson3}} presumably does not avoid this concern; \ the same issue
could arise in their scenario if one imagines trying to violate the GSL with a
continuous flux of particles undergoing the classical process they describe.
Even if one takes their background to be given by, for example,
Einstein-Aether theory, the presence of a flux of particles coupled to this
background could result in it being slightly shifted.  Such a shift might
lead to a sufficient flow of positive energy into the black hole to save the GSL.}}

%\ Ignoring parametric
%dependence on the velocities, the energy flux at the horizon for the type $1$
%or $2$ particles is roughly of order $T^4$. \ These fluxes perturb the ghost
%condensate by a small amount; it follows from the form of the coupling that
%the induced perturbation to $\varphi$ is suppressed at the horizon by roughly
%a factor $T^4 / \Lambda^4$. \ This is smaller than the effect we were worried
%about from the higher derivative operators, but it could still cause a
%problem; \ we might worry that we could obtain $T^r_t$ as large as $T^4$ for the ghost
%condensate at the horizon, which is comparable to the flux from Hawking
%radiation. \ We thus cannot be sure if the perpetual motion machine is ruined
%without a more careful analysis.

\paragraph{}It should be clear, therefore, that a more thorough examination of
the energy flow in the ghost condensate during the running of the perpetual
motion machine of {\cite{Sergeis}} is crucial to determine whether or not the
GSL is actually violated. \ For this reason, we will turn to consider the
nature of such flows in more detail in the following section.  There we will find
that the perpetual motion machines can in fact be made to work, but that the
reasons for this are closely related to the existence of a set of negative
energy ghost condensate flows.  These may then be use to violate the
generalized second law in a very simple way.

\section{Ghost Condensate Flows\label{ghost}}

Note that we have been making an important implicit assumption in our
discussion of ghost condensate flows thus far; that the stationary rate of
energy flow in the ghost condensate into a black hole is actually a uniquely
defined quantity. \ This assumption stems from an important result in fluid
dynamics: \ Given a perfect fluid surrounding a black hole, and a particular
asymptotic density, there exists a unique stationary spherically symmetric
flow of the fluid into the hole (see {\cite{astro}} and references therein). \
Let us briefly review the reason for this:

\paragraph{}Consider a perfect fluid with a density $\rho$, pressure P, four
velocity $u^{\mu}$, and \ ``baryon'' number density $n$. \ The speed of sound,
which we will take to be subluminal for simplicity, is given by $a =
\sqrt{\frac{\mathd P}{\mathd \rho}}$, and the energy momentum tensor is
\begin{equation}
  T_{\mu \nu} = (\rho + P) u_{\mu} u_{\nu} - g_{\mu \nu} P \label{fluidT} .
\end{equation}
Assuming the black hole is much heavier than the energy of fluid flowing into
it (so that we may ignore back-reaction effects), the key equations describing
the stationary, spherically symmetric flow of the fluid are:
\begin{equation}
  \tmop{baryon} \tmop{conservation} : \nabla_{\mu} (n u^{\mu}) = 0 \Rightarrow
  \partial_r (n u^r r^2) = 0 .
\end{equation}
\begin{equation}
  \tmop{adiabaticity} : d ( \frac{\rho}{n}) + P d ( \frac{1}{n}) = 0
  \Rightarrow \frac{d \rho}{d n} = \frac{\rho + P}{n} .
\end{equation}
\begin{equation}
  \tmop{energy} \tmop{conservation} : \nabla_{\mu} T^{\mu}_t = 0 \Rightarrow
  \partial_r \left( (\rho + P) u^r u_t r^2 \right) = 0.
\end{equation}
Using these relations, as well as the fact that $u_{\mu} u^{\mu} = 1$, we may
solve for the radial derivative of $u \equiv |u^r |$, obtaining
\begin{equation}
  \frac{d u}{d r} = \frac{N}{D}, \label{uprime}
\end{equation}
where
\begin{equation}
  N = (1 - 2 M / r + u^2) \frac{2 a^2}{r} - \frac{M}{r^2},
\end{equation}
and
\begin{equation}
  D = \frac{u^2 - (1 - 2 M / r + u^2) a^2}{u} . \label{D}
\end{equation}
\paragraph{}Since we assume the fluid to be at rest far from the black hole,
we know that at large $r$, $u$ approaches zero. \ In particular, the flow at
large $r$ becomes sub-sonic with $u^2 \ll a^2$, which in turn implies that $D
< 0$. \ Close to the horizon on the other hand, inspection of (\ref{D})
reveals that $D > 0$. \ We may therefore conclude that at {\tmem{some}} point
outside the horizon, $D$ passes identically through zero. In fact, it turns
out that at this location the flow velocity as measured by a stationary
observer is exactly passing through the speed of sound- it is the location of
the sound horizon. \ In order for the flow to successfully pass through the
sound horizon without becoming singular, it is clear that equation
(\ref{uprime}) requires that, at this radius, $N = 0.$ \ This extra condition,
required to ensure smoothness of the flow at the sound horizon, now provides
us with enough information to solve for the flow exactly.

\paragraph{}Naively, we have a very good reason to believe that this argument
should apply equally well to the case of the ghost condensate. \ In fact, the
ghost condensate may in a precise sense be thought of as a degenerate case of
a perfect fluid {\cite{ghost2}}: Suppose we consider any Lagrangian of the
form
\begin{equation}
  \mathcal{L}= P (X) \label{PofX},
\end{equation}
with $X = \nabla_{\mu} \varphi \nabla^{\mu} \varphi$ as before. \ The ghost
condensate is of course described by a special case of such a
Lagrangian.{\footnote{In fact, the ghost condensate is generally defined by a
point $X_0$ such that $X_0 > 0$, $P' (X_0) = 0$, and $P'' (X_0) > 0$. \ The
specific form of the Lagrangian in equation (\ref{L}) was taken only for
simplicity.}} \ The energy momentum tensor associated with (\ref{PofX}) is
then
\begin{equation}
  T_{\mu \nu} = 2 P' (X) \nabla_{\mu} \varphi \nabla_{\nu} \varphi - g_{\mu
  \nu} P (X) . \label{XT}
\end{equation}
The key point is then that this system describes precisely a perfect fluid, so
long as we make a \ set of identifications:
\begin{eqnarray}
  P (X) \longrightarrow P &  &  \nonumber\\
  \nabla^{\mu} \varphi / \sqrt{X} \longrightarrow u^{\mu} &  &  \nonumber\\
  2 P' (X) X - P (X) \longrightarrow \rho &  &  \nonumber\\
  \frac{P' (X)}{2 P'' (X) X + P' (X)} \longrightarrow a^2 . &  & 
\end{eqnarray}
Additionally, $n$ is taken proportional to $P' (X) \sqrt{X}$ so that the
equation of motion for $\varphi$ becomes the statement of baryon number
conservation.

\paragraph{}We might thus expect that ghost condensate flows into black holes
should be unique, exactly as in the perfect fluid case. \ On the other hand, a
closer inspection of the argument reveals at least a conceivable loop-hole: \
The speed of sound in the ghost condensate, in the absence of higher
derivative terms, is equal to zero (since $P' (X) = 0$). \ Thus a ghost
condensate flow, at the level of the perfect fluid picture, never actually
passes through a sound horizon. \ In fact, as we will now show, this loop-hole
does indeed allow for the existence of a family of stationary, spherically
symmetric ghost condensate flows.

\paragraph{}Let us look for well behaved solutions to the ghost condensate
equation of motion in the Schwarzschild background which are small
perturbations to the zero-energy solution (\ref{phi}). \ Again we will look
for solutions which are stationary and spherically symmetric, and we will
ignore back-reaction of the flow on the background metric. \ Stationarity
requires that the components of $\nabla^{\mu} \varphi$, in Schwarzschild
coordinates (\ref{metric2}), should be independent of the time $t$. \ This in
turn requires that $\varphi$ must take the form $\varphi = A t + f (r)$, for
some constant $A$ and function $f$. \ Since we want the physics of the
solution to approach that of the usual flat space configuration $\varphi =
\Lambda^2 t$ at large distances from the black hole, we must continue to take
$A = \Lambda^2$. \ The general perturbation of interest to the zero-energy
solution (\ref{phi}) thus takes the form
\begin{equation}
  \varphi = \Lambda^2 \tau + \pi (r) \label{perturbed} .
\end{equation}
Plugging this into the equation of motion (\ref{curvedpi}), we obtain an
extremely simple differential equation
\begin{equation}
  \frac{2 M}{r} \pi'' (r) + \frac{2 M}{r^2} \pi' (r) = 0, \label{linearpi}
\end{equation}
with a simple set of solutions
\begin{equation}
  \pi (r) = C_1 \log (r) + C_2 . \label{pisol}
\end{equation}
The constant solution $C_2$ was to be expected, since our theory has a shift
symmetry $\pi \rightarrow \pi + c$, but the existence of the $\log$ solutions
is surprising; \ they are perfectly well behaved through the event horizon out
to infinity, and as such represent a one parameter family of stationary,
spherically symmetric ghost condensate flows. \ Note that the fact that $\log
(r)$ blows up as $r \rightarrow \infty$ is irrelevant, as the shift symmetry
ensures that only derivatives of $\pi$ are physical. \ Indeed, let us check
for what values of $C_1$ (\ref{pisol}) actually represents a small
perturbation about the zero-energy solution (\ref{phi}). \ Expanding $X$ to
linear order in the perturbation, we obtain
\begin{equation}
  X \simeq \Lambda^4 - 2 C_1 \frac{\Lambda^2}{r} \sqrt{\frac{2 M}{r}} .
  \label{Xperturb}
\end{equation}
As long as we are only interested in values of $r$ of order the horizon size
or larger, we conclude that our perturbative solutions are valid if we take
\begin{equation}
  |C_1 | \ll \frac{\Lambda^2}{T} . \text{} \label{Cineq1}
\end{equation}
\paragraph{}Another thing we should check is that these flow solutions are not
significantly affected by the presence of higher derivative terms in the
Lagrangian such as (\ref{higherd}). In fact, as discussed in section
(\ref{GSL}), it is known that higher derivative terms have a small effect on
the original zero-energy solution (\ref{phi}), suppressed by the small
quantity $T^2 / \Lambda^2$. \ More quantitatively, the leading order
perturbation $\lambda$ induced by these terms is a solution to the linear
equation (\ref{linearpi}), but with a source given by the appropriate higher
derivative term evaluated on the zeroth order background. \ With the term from
(\ref{higherd}) we then have
\begin{equation}
  \frac{2 M}{r} \lambda'' (r) + \frac{2 M}{r^2} \lambda' (r) = -
  \frac{\alpha}{\Lambda^2} \Box^2 \varphi_0 = - \alpha \frac{9 (r - 6 M)}{4
  \sqrt{2} r^4} \sqrt{\frac{M}{r}} .
\end{equation}
Setting a boundary condition at infinity so that $\lambda$ does not affect the
radial energy flux, we obtain the solution
\begin{equation}
  \lambda = - \alpha \frac{3 (3 r - 2 M)}{2 r^2 \sqrt{2 M / r}} .
\end{equation}
Comparing with the size of the perturbation defined by our flow solution
(\ref{pisol}), we may conclude that the perturbation coming from the higher
derivative term is a sub-dominant effect on scales of horizon size or larger
so long as we take
\begin{equation}
  |C_1 | \gg T. \label{Cineq2}
\end{equation}
Note that since the flows (\ref{pisol}) with non-zero $C_1$ are perfectly well
behaved perturbations to the zeroth order background (\ref{phi}), they
themselves change the effect of the higher derivative terms by an amount which
is small so long as (\ref{Cineq1}) is satisfied. \ In combination, the
inequalities (\ref{Cineq1}) and (\ref{Cineq2}) thus leave us with a large
range of sizes for these flows for which we may trust the form of our
perturbative solutions, while ignoring the effects of higher derivative
operators.

\paragraph{}Before we move on to study the ghost condensate black hole flows
in more detail, we'd like to point out an interesting observation
concerning the general perfect fluid flows discussed earlier. \ Specifically,
suppose we have a Lagrangian of the type (\ref{PofX}), with a non-zero speed
of sound at large distances, and suppose we add to it some small perturbation.
\ The perturbation could be for example a higher derivative operator, or a
coupling to some secondary field, etc. \ The uniquely defined flow into a
black hole will of course be changed by some small amount. \ In particular, it
turns out that in the expression (\ref{uprime}) for $\frac{d u}{d r}$, the
denominator ``$D$'' is unaffected, while the numerator ``$N$'' is shifted by a
term small in the perturbation added to the action. \ It thus follows that the
argument for uniqueness of the flow goes through unaffected- $D = 0$ still
implies $N = 0$ in order to avoid singularities. \ On the other hand, the
parameters determining the flow, as set by the $N = 0$ requirement, are
changed by an amount given by the \ size of the perturbation as
evaluated at the sound horizon.  If the ghost condensate black hole flows
were unique, this argument would tell us how they would become affected by
small perturbations to the action such as those considered in section (\ref{GSL}).
\ Of course, since the uniqueness argument actually breaks down for the case
of the ghost condensate, this point is moot.

\paragraph{}In the next section, we will discuss the form of the
energy-momentum tensor for the family (\ref{pisol}) of ghost condensate black
hole flows. \ We will show that for appropriate $C_1$ they carry negative
energy, and we will also make some comments concerning their stability.

\section{Energy and Stability\label{energy}}

Expanding the energy-momentum tensor (\ref{T}) to linear order about the $X =
\Lambda^4$ background, we obtain
\begin{equation}
  T^{\mu \nu} \simeq 4 \Lambda^2 \xi^{\mu} \xi^{\nu} \xi^{\alpha}
  \nabla_{\alpha} \pi . \label{linearT}
\end{equation}
With the specific form of the solutions (\ref{pisol}) this yields
\begin{equation}
  T^r_t \simeq \frac{8 C_1 M \Lambda^2}{r^2} \label{T0r}
\end{equation}
and{\footnote{Note that the singularity in $T^t_t$ at the horizon is not
physical, but due to the Schwarzschild coordinate singularity. \ Freely
falling observers measure a perfectly finite energy at the horizon.}}
\begin{equation}
  T^t_t \simeq - \frac{4 C_1}{r} \sqrt{\frac{2 M}{r}} \frac{\Lambda^2}{1 - 2 M
  / r} . \label{T00}
\end{equation}
We thus find that, depending on the sign of $C_1$, these flow solutions
describe either regions of positive energy with positive energy flux into the
black hole, or regions of {\tmem{negative energy}} with a corresponding
{\tmem{{\tmem{negative energy flux}}}}. \ Such flows therefore allow one to
trivially violate the generalized second law of thermodynamics in this theory;
\ they cause the horizon area to decrease without any corresponding increase
in the total entropy of the exterior.{\footnote{We are assuming that the
    ghost condensate flows are accompanied by positive fluxes of entropy into
  the black hole.  If the UV completion of the theory were such that large
  outward entropy fluxes were associated with these flows, then saving the
  generalized second law might become possible.  This could conceivably
  occur, for example, if the UV completion of the theory contained particles
  with faster than light propagation.  Note that Hawking radiation generates
  entropy in the exterior at a rate which is far too small to be relevant
  given the size of the negative energy flows under consideration (c.f. (\ref{Cineq1})
  and (\ref{T0r})).  Hawking radiation of ghost condensate quanta themselves has been
  shown to be especially suppressed {\cite{feldstein}}.}} \ This demonstrates in particular that
the ghost condensate is incompatible with the GSL independent of the size or
nature of couplings to secondary fields. \ It had been argued, for example,
that GSL violation might be avoided if direct couplings between the ghost
condensate and other fields were induced only \ gravitationally
{\cite{shinji2}}. \ Moreover, this shows that perpetual motion machines such
as those of {\cite{Sergeis,Jacobson3}} may indeed be constructed; there will
certainly be flows for which the ghost condensate energy accretion rate is
sufficiently small.

\paragraph{}Note that, if we take these ghost condensate solutions literally
at extremely large distances from the black hole, then our approximation of
ignoring the back-reaction on the metric will become a bad one: \ $T^t_t$ from
$_{}$(\ref{T00}) falls off at infinity like $1 / r^{3 / 2}$, and thus the
total integrated energy is not finite.{\footnote{This explains why these
solutions were not identified in {\cite{shinji}}, since in that treatment the
metric was assumed to be asymptotically flat.}} \ Of course, any stationary
fluid flow solution into a black hole with a non-zero asymptotic density will
have this same problem. \ The point is that one should think of these flows as
being approximately stationary for a very long period of time, before the
(appropriately finite) total energy available has been exhausted. \ This time
period can be made arbitrarily longer than the Schwarzschild time-scale by
raising $M_{\tmop{pl}}$ (for fixed $r_h$), thereby increasing the mass of the
black-hole and reducing the back-reaction.

\paragraph{}Another important issue for these negative energy flows is that of
stability. \ Recall that in flat space, ghost condensate configurations with
$X = \Lambda^4$ have perturbations which lack a $\vec{\nabla}^2 \pi$ term in
their wave equations. \ An important point though, is that if we consider
configurations with $X < \Lambda^4$ instead, say $X = \Lambda^4 - \Delta$, a
$\vec{\nabla}^2 \pi$ term does appear, but with the wrong sign:
\begin{equation}
  (1 - \frac{3 \Delta}{2 \Lambda^4}) \ddot{\pi} \label{unstable2} = -
  \frac{\Delta}{2 \Lambda^4}  \vec{\nabla}^2 \pi - \frac{\alpha}{\Lambda^2}
  \vec{\nabla}^4 \pi .
\end{equation}
Such background configurations are of course unstable, and in fact correspond
to the negative energy region of the theory. \ In addition, it follows that
there exists a {\tmem{non-linear}} instability in the original $X = \Lambda^4$
background {\cite{ghost2}}; a short wavelength mode living on top of a longer
wavelength mode with $X$ values temporarily in the negative energy region will
indeed display the instability.

\paragraph{}It thus seems clear that the negative energy flows we have
constructed will be unstable. \ Although analytic solutions for the growing
modes have not been forthcoming, we can still attempt to make a rough estimate
for the time-scale of their growth. \ Note that for a given value of $\Delta$,
higher derivative operators put an upper limit on the wavenumber for an
instability. \ In particular, in order for the $\vec{\nabla}^2 \pi$ term in
(\ref{unstable2}) to dominate over the stabilizing higher derivative term, we
require the wavenumber $k$ to be smaller than roughly $\sqrt{\Delta} /
\Lambda$. \ The corresponding timescale for the growth of the mode, from
(\ref{unstable2}), is then of order $\Lambda^3 / \Delta$.

\paragraph{}In the ghost condensate flow solutions we have been considering,
$\Delta X$ outside the horizon has a maximum magnitude of roughly $C_1
\Lambda^2 T$, corresponding to a decay timescale of order $\Lambda / C_1 T$. \
Given the energy flux from equation (\ref{T0r}), the total energy which may be
transferred in this time is of order $\Lambda^3 / T^2$. \ The corresponding
change in the black hole area in Planck units is then of order $\Lambda^3 /
T^3$, and thus one may easily violate the generalized second law before the
instability of these flows manifests itself.{\footnote{To be more precise, we
could take $C_1 \gg \Lambda$ for simplicity, so that the evolution of the
unstable modes takes place on scales much smaller than $r_h$. \ We could then
roughly trust the form of the flat space wave-equation, and $\Lambda / C_1 T$
would be the approximate instability timescale as seen by a freely falling
observer at rest relative to the ghost condensate. \ If we set up the initial
conditions for the flow on a well behaved space-like hypersurface such as
$\tau = 0$, for example, then the instability will become important at
$\tau$'s of order $\Lambda / C_1 T$. \ At a fixed radius $r$, it then follows
from (\ref{tau}) that the killing time $t$ available is also given roughly by
$\Lambda / C_1 T$, so that the total energy transferred before the flow
becomes affected by the instability will be approximately $\Lambda^3 / T^2$ as
claimed. \ Flows with smaller values of $C_1$ may be used to violate the GSL
on longer timescales, although precise general relativistic values for the
lifetimes then become more difficult to estimate.}}

\paragraph{}In fact, the nature of the instability of negative energy regions
in the ghost condensate was studied numerically in {\cite{ghost2}}. \ Those
authors found that negative energy lumps in the ghost condensate tend to
shrink in size (rather than grow catastrophically), while maintaining a fixed
total amount of negative energy. \ The fact that the tendency of these
negative energy lumps is to coalesce suggests the the nature of the
instability in the black hole flows will not be a particularly remarkable one;
presumably, the flows will very gradually accelerate, until eventually all of
the negative energy around the black hole has been absorbed.

\section{Discussion\label{discussion}}

In this paper we have demonstrated the existence of a one parameter family of
stationary, spherically symmetric solutions describing flow of ghost
condensate fluid into black holes. \ These flows may carry either positive or
negative energy and may thus be used to violate the generalized second law of
thermodynamics.

\paragraph{}One point that is slightly puzzling about the existence of these
flows is related to what happens if one perturbs slightly away from a pure
ghost condensate solution at large distances from the black hole. \ In
particular, suppose that as $r \rightarrow \infty$, $X$ approaches $\Lambda^4
(1 + \delta)$, so that the field approaches a configuration slightly different
than the ghost condensate one at $X = \Lambda^4$. \ In this case the standard
argument of section (\ref{ghost}) for uniqueness of perfect fluid flows into
black holes applies, and thus the one parameter family of solutions must
disappear. \ Presumably what is happening is that for non-zero $\delta$, there
is indeed a unique stationary flow, but also there is a family of flows with
some small time dependence set by the size of $\delta$. \ As $\delta
\rightarrow 0$, the time dependence goes away, and we recover the complete
family of stationary ghost condensate flows. \ It would certainly be
interesting to see this explicitly, though this is beyond the scope of the
present paper.

\paragraph{}The fact that it is possible to violate the GSL through negative
energy flow in the ghost condensate suggests that it is really violation of
the null energy condition (NEC) which is the source of the problems this
theory causes for thermodynamics. \ In particular, as we have discussed, the
perpetual motion machines of {\cite{Sergeis,Jacobson3}} hinge fundamentally on
the requirement that the background violating Lorentz invariance also carries
no energy-momentum. \ Generically, the Lorentz violating background field can
then be expected to have perturbations which contribute to $T_{\mu \nu}$ at
linear order as in (\ref{linearT}), leading directly to violations of the NEC.

\paragraph{}It thus seems likely that it is not so much Lorentz violation
which leads to potential problems with thermodynamics, as implied by the
perpetual motion machines of {\cite{Sergeis,Jacobson3}}, but simply the
existence of negative energy configurations. \ Of course, the test of this
hypothesis comes in trying to find a Lorentz violating theory for which the
NEC is satisfied, but for which one can successfully set up a perpetual motion
machine.

\paragraph{}The first thing one might try in this direction would be to
continue to consider a general Lagrangian of the form $\mathcal{L}= P (X)$ as
in section (\ref{ghost}). \ One could then demand the requirements that not
only is there a point $X_0$ with $P' (X_0) = 0$ as in the ghost condensate (to
allow a background with zero energy), but that also $P'' (X_0) = 0$ in order
to remove the linear term (\ref{linearT}) in $T_{\mu \nu}$ coming from
fluctuations about $X_0$. \ The problem here though, is that the action for
these perturbations then begins at cubic order, and \ the theory then becomes
completely non-perturbative. \ As a result, conclusions drawn in such a
scenario could not be trusted.

\paragraph{}Another interesting example to consider is that of Einstein-Aether
theory {\cite{Aether0,Aether1,Aether2,BlackAether,EnergyAether}}. \ This
theory possesses a vector field with a time-like expectation value, and is
known to have choices of parameters for which energies are apparently
non-negative {\cite{EnergyAether}}. \ There are also parameter values for
which black hole solutions have been studied {\cite{BlackAether}}, and it
would be very interesting to investigate whether one can construct a perpetual
motion machine in this theory without sacrificing the positive energy
requirements.  \ A natural conjecture would be that it will not in fact be
possible.  To check this one would have to study, in particular, whether one could
maintain a sufficiently small energy flux in the Einstein-Aether fields into the
black hole in the presence of the ``slow'' and ``fast'' particles being used
to run the perpetual motion machine.

\paragraph{}The fact that the pure ghost condensate theory is incompatible
with the generalized second law of thermodynamics fits in nicely with the
result of {\cite{feldstein}}, that the Hawking spectrum of ghost condensate
perturbations is highly suppressed and non-thermal. \ The obvious suggestion
would be that the ghost condensate cannot emerge in a low energy effective
theory coming from a quantum theory of gravity.{\footnote{Of course it is
always possible that quantum gravity will not ultimately behave as we
expect.}}

\paragraph{}The ghost condensate has been proposed as a concrete model for
both dark energy and dark matter, as well as inflation. \ It is quite
interesting that a seemingly consistent model describing low energy phenomena
far below the Planck scale could turn out to be unacceptable due to the
behavior of black holes as required by fundamental principles of gravity.

\section{Acknowledgements}

The author would like to thank Sergei Dubovsky and Ben Freivogel for useful
discussions. \ This work was supported by the Department of Energy grant
number DE-FG02-01ER-40676.

\end{document}